\documentclass{article}%%{{{1
\usepackage{amsmath}
\usepackage{amssymb}
\allowdisplaybreaks[1]
\usepackage[numbers,sort&compress]{natbib}
\usepackage{color}
\usepackage[colorlinks=true,urlcolor=blue,citecolor=blue,linkcolor=blue]{hyperref}
\usepackage[textwidth=0.72\paperwidth,textheight=0.76\paperheight]{geometry}
\usepackage{booktabs}
\usepackage{mathfunc}
\def\mathindent{25pt}
\makeatletter
\def\PK{\mathop{\operator@font PK}\nolimits}
\def\DK{\mathop{\operator@font DK}\nolimits}
\makeatother

\begin{document}

\title{Analytic calculation of doubly heavy hadron spectral density\\
in coordinate space}
\author{Zhi-Wei Huang and Jueping Liu\footnote{Corresponding author.}\\
College of Physics and Technology, Wuhan University, Wuhan 430072, China\\
(E-mail: \texttt{zwhuang@whu.edu.cn}, \texttt{jpliu@whu.edu.cn})}

\maketitle

\begin{abstract}
A systematic and easy-to-use method is developed to
calculate directly the doubly heavy hadron spectral density in the coordinate space.
The correlation function is expressed in terms of hypergeometric functions, and the
spectral density is obtained through two independent approaches: the simple
integral representation method and the epsilon-expansion method,
respectively. It is found that the spectral density of doubly heavy hadrons
can be analytically expressed through commonly known simple functions.
This method can drastically simplify and improve the QCD spectral sum rule
calculation of the doubly heavy hadrons.
An instructive numerical method is also presented for fast evaluation of the
spectral density.
\end{abstract}


%%}}}1
%%============================================================================
\section{Introduction}%%{{{1

In the past years, lots of $XYZ$ states
\cite{Swanson:2006st,Klempt:2007cp,Brambilla:2010cs} have been
observed by BaBar and Belle collaborations. Many of these states seem not to
have a conventional $c\bar{c}$ or $b\bar{b}$ structure, which inspired the extensive
study of the exotic hadron spectroscopy. Then many possible structures
like tetraquark, hybrid or molecular state are proposed and studied
\cite{Matheus:2006xi,Navarra:2007yw,Matheus:2009vq,Narison:2010pd,
Wang:2009bd,Wang:2010uf, Chen:2010ze,Du:2012pn, Berg:2012gd} with QCD spectral
sum rules \cite{Shifman:1978bx,Shifman:1978by, Reinders:1984sr}.

These structures can be formally expressed as $\{QQ'X\}$, where $Q$ ($Q'$) denotes
heavy quark or antiquark and $X$ denotes light quarks and/or gluons. The QCD spectral sum
rules require the knowledge of the two-point correlation function or its
discontinuity, the spectral density, to study many fundamental properties of
hadrons. Spectral density is the most labor-intensive part of the QCD spectral
sum rule calculation.
Therefore, it is important and helpful to develop a systematic and easy-to-use
method to calculate doubly heavy hadron spectral density.

For a hadron state of light quarks and/or gluons, the correlation function
can be easily calculated by a Fourier transformation in the coordinate space
\cite{Yang:1993bp,Wang:2008vg}, because the light quark/gluon masses are
relatively small and the propagators can be approximated by fractional
functions. Once the heavy quarks are introduced into this kind of states,
one need either to use the momentum space representation of the heavy quark
propagators to keep the masses finite and then evaluate momentum integrals
\cite{Kim:2004pu,Matheus:2006xi}, or to handle Bessel function related
integrals in coordinate space \cite{Groote:2005ay}.

In recent years, the momentum space method \cite{Kim:2004pu,Matheus:2006xi}
has been widely used for mesons
\cite{Matheus:2006xi,Navarra:2007yw,Matheus:2009vq,Nielsen:2009uh,
Zhang:2009vs,Zhang:2011jja, Chen:2010jd, Wang:2009bd,Wang:2010pn,
Albuquerque:2012rq} and baryons \cite{Zhang:2008pm} at the leading order in
$\alpha_s$. Even so, it is tedious to evaluate two-loop momentum integrals
with the traditional methods, especially for integrals with tensor structures.
What's more, the double integral representations are time-consuming and might
introduce noticeable errors to the sum rules. 

In this work, a direct and simple method of analytically calculating the
doubly heavy hadron spectral density in the coordinate space is developed. In
this method, the heavy
quark propagator is expressed in the form of the modified Bessel function of the
second kind, and the correlation function can be expressed by a few generalized
hypergeometric functions. The hypergeometric representations make the
correlation function easily expressed in a compact form and the spectral
density easy to be worked out analytically. Note that the negative dimensional
integration method (NDIM) \cite{Suzuki:1997yz,Suzuki:1998qv,
Anastasiou:1999ui} or its optimized version, the method of brackets (MB)
\cite{Gonzalez:2007ry,Gonzalez:2008xm,Gonzalez:2011nq,Gonzalez:2010nm}, is
used to calculate some important integrals.

In Refs.~\cite{Broadhurst:1993mw,Groote:2005ay,Greynat:2011zp}, the spectral
density is calculated in several special cases, where small
momentum expansion $q^2\to0$, large momentum expansion $q^2\to-\infty$ and
threshold expansion are used to simplify the calculation. Obviously, the
spectral density calculated in these special cases is not valid for a wide
energy region of the QCD spectral sum rules calculation. 

In this work, the spectral density that valid for a wide energy region is
calculated in two independent approaches: the simple integral representation
method and the $\epsilon$-expansion method, respectively. The simple (onefold)
integral representation \cite{Broadhurst:1993mw} of $\HypPFQ{q+1}{q}$ type
hypergeometric functions is suitable for spectral density calculation, where
the knowledge of the power function's discontinuity is enough to obtain the
result. In the recent
decade, lots of algorithms or packages \cite{Moch:2001zr,Weinzierl:2002hv,
Huber:2005yg,Huber:2007dx, Kalmykov:2006hu,Kalmykov:2007pf, Bytev:2011ks} have
been developed to perform the $\epsilon$-expansion of hypergeometric
functions. Practically, \verb|HypExp|~\cite{Huber:2005yg,Huber:2007dx} is used to
perform $\epsilon$-expansion of $\HypPFQ{q+1}{q}$ type hypergeometric
functions. It is found that the spectral density of doubly heavy hadrons can
be analytically expressed in terms of commonly known simple functions and no parameter
integral is needed at all.

The well regularized hypergeometric representation of the spectral density can
be evaluated numerically, too. An instructive method is developed for
numerical computation of the spectral density. In this method, 
no analytic $\epsilon$-expansion is needed, and one can treat hypergeometric
functions as common functions. Consequently, the spectral density can be
computed directly, which will be helpful to the standard community.

Generally, the coordinate space method can be widely applied to the QCD
spectral sum rule calculations of doubly heavy hadron states $\{QQ'X\}$. This method
provides a systematic and easy-to-use approach to calculate directly the
doubly heavy hadron spectral density in the coordinate space, which
drastically simplifies the QCD spectral sum rule calculation. Expressing the
spectral density in terms of simply functions makes the QCD sum rule
calculation extremely efficient. There is no noticeable errors from
multi-dimensional numerical integrals, either. Then a Monte-Carlo based
uncertainty analysis \cite{Leinweber:1995fn,Wang:2011zzx} is feasible to make
realistic uncertainty estimates of the phenomenological parameters.
As the tetraquark or molecular structure of these states leads to (almost) the
same predictions within the accuracy \cite{Narison:2010pd},
such analysis can quantitatively improve the predictive ability of QCD sum
rules. Moreover, this method can also serve to be an important cross-check of
the widely used momentum representation method.

The paper is organized as follows. After a brief introduction of the
traditional momentum space method in the next section, the method of
calculating the doubly heavy hadron spectral density in the coordinate space
is presented in detail in Sec.~\ref{sec:coord}. The compact hypergeometric
representation of the correlation function is derived in
Sec.~\ref{sec:hyprep}. Then two approaches to extract discontinuities from the
hypergeometric functions are provided in Sec.~\ref{sec:intrep} and
Sec.~\ref{sec:hypexp}, respectively. In Sec.~\ref{sec:egapp}, a concrete
calculation is performed to show how to apply the coordinate space method to
the practical problems.
In Sec.~\ref{sec:numhyp}, an instructive numerical method is also presented for
fast evaluation of the spectral density from the $\epsilon$-regularized
hypergeometric functions.
Finally, a summary is given in Sec.~\ref{sec:con}.

%%}}}1
%%============================================================================
\section{The momentum space method}%%{{{1
\label{sec:mom}

The QCD spectral sum rules use the two-point correlation functions like
\begin{equation}
\Pi_{\mu\nu}(q)
=i\int d^4x\,e^{iq.x}\langle0|T\{j_\mu(x)j_\nu^\dag(0)\}|0\rangle
\label{eq:pimunu}
\end{equation}
to study lots of properties of hadrons. Generally, the correlator in
coordinate space can be expressed as a product of full propagators and/or
their derivatives \cite{Groote:2005ay}. And then the correlation function in
the momentum space can be expressed as a Fourier transformation of the
product of several propagators. For example, the Lorentz invariant two-point
correlation function of $\{QQqq\}$ state in coordinate space is of the simple
form
\begin{equation}
\Pi(q^2)\sim i\int d^Dx\,e^{iq.x}S_Q(x)S_Q(-x)S_q(x)S_q(-x) ,
\label{eq:piprops}
\end{equation}
where all color indices, flavor indices and gamma matrices are ignored, for the
sake of briefness. The discontinuity of $\Pi(q^2)$ is the spectral density,
which is needed by the QCD spectral sum rules calculation.

In the past years, the discontinuity of this type of integrals is calculated
for mesons \cite{Matheus:2006xi,Navarra:2007yw,Matheus:2009vq,Nielsen:2009uh,
Zhang:2009vs,Zhang:2011jja, Chen:2010jd, Wang:2009bd,Wang:2010pn,
Albuquerque:2012rq} and baryons \cite{Zhang:2008pm} at the leading order in
$\alpha_s$ by using the techniques in Ref.~\cite{Kim:2004pu,Matheus:2006xi}.
To keep the heavy quark mass finite, the momentum space expression of the
heavy quark propagator
\cite{Reinders:1984sr}
\begin{align}
iS_Q^{ab}(p)=&
  \frac{i\delta^{ab}}{\hat{p}-m}
  +\frac{i}{4}\frac{\lambda_{ab}^{n}}{2}g_s G_{\mu\nu}^{n}
   \frac{\sigma^{\mu\nu}(\hat{p}+m)+(\hat{p}+m)\sigma^{\mu\nu}}
        {(p^2-m^2)^2} \cr
 &
  +\frac{i\delta^{ab}}{12}\langle g_s^2 G^2\rangle m
   \frac{p^2+m\hat{p}}{(p^2-m^2)^4}
  +\cdots 
  \label{eq:prophQmom}
\end{align}
is used, where $\hat{p}=\gamma_\mu p^\mu$. The light quark part of the
correlation function is calculated in the coordinate space, and then the whole
expression is converted to the momentum space by a $D$ dimensional Fourier
transformation. Finally, the two-loop momentum integral is calculated and the
spectral density is expressed as double integrals of Feynman or Schwinger
parameters.

The two-loop momentum integral is of the form
\begin{equation}
\int\frac{d^4k_1 d^4k_2}{(2\pi)^8}
\frac{\mathcal{N}(k_1,k_2,q,m)}
     {[-k_1^2+m^2]^{n_1}[-k_2^2+m^2]^{n_2}[-(q-k_1+k_2)^2]^{\nu_3}} .
\label{eq:mom2loop}
\end{equation}
It is tedious to calculate this type of integrals in the traditional manner,
and the non-trivial numerator $\mathcal{N}(k_1,k_2,q,m)$ makes the calculation
even complicated. Besides, the double integral representations of Feynman or
Schwinger parameters are time-consuming and might introduce noticeable errors
to the sum rules. One can also use the Mellin-Barnes method to express such
integrals as combinations of hypergeometric functions, but the tensor
structure makes the final expression diffuse. Therefore, it is of great
significance to find a convenient way of calculating the spectral density.

Actually, the spectral densities of doubly heavy hadrons can be calculated in
a straightforward way in the coordinate space. In the following section, a
direct and simple method is developed to calculate the doubly heavy spectral
density in the coordinate space.

%%}}}1
%%============================================================================
\section{The coordinate space method}%%{{{1
\label{sec:coord}

Since the correlation function can be expressed as the Fourier transformation of a
product of propagators in the coordinate space, it is natural to calculate the
correlation function in the coordinate space. The light quark propagator and
the gluon propagator can be expressed as common functions, because the masses
are small and the small mass expansion can be used. As for the heavy quark,
the modified Bessel function of the second kind $K_n(mx)$ is needed to express
the propagator in the coordinate space. Note that the so-called fixed-point
gauge ($x^\mu A_\mu(x)=0$) technique is used to get the full propagators.

%%============================================================================
\subsection{Propagators}%%{{{2
\label{sec:propagator}

The free-standing part of the light quark propagator in the coordinate space is
\cite{Yang:1993bp, Wang:2008vg,Wang:2011zzx}
\begin{align}
iS_q^{ab}(x)
=&\frac{\delta^{ab}}{2\pi^2 x^4}\widetilde{x}
  -\frac{\delta^{ab}}{12}\langle\bar{q}q\rangle
  -\frac{\delta^{ab}g_s^2 x^2\widetilde{x}}{2^5\times 3^5}\langle\bar{q}q\rangle^2 \cr
 &
  +\frac{1}{32\pi^2} g_s G_{\mu\nu}^{ab}
   \frac{\sigma^{\mu\nu}\widetilde{x}+\widetilde{x}\sigma^{\mu\nu}}{x^2}
  +\frac{\delta^{ab}x^2}{192}\langle g_s\bar{q}\sigma Gq\rangle \cr
 &
  -\frac{\delta^{ab} x^4}{2^{10}\times 3^3}\langle\bar{q}q\rangle\langle g_s^2 G^2\rangle
  +\cdots \cr
 &
  -\frac{m_q\delta^{ab}}{4\pi^2 x^2}
  +\frac{m_q\delta^{ab}\widetilde{x}}{48}\langle\bar{q}q\rangle
  -\frac{m_q \delta^{ab} g_s^2 x^4}{2^7\times 3^5}\langle\bar{q}q\rangle^2 \cr
 &
  +\frac{m_q}{32\pi^2} g_s G_{\mu\nu}^{ab}\sigma^{\mu\nu}\ln(-x^2)
  -\frac{m_q\delta^{ab}x^2\widetilde{x}}{2^7\times 3^2}\langle g_s\bar{q}\sigma Gq\rangle \cr
 &
  -\frac{m_q\delta^{ab}}{2^9\times 3\pi^2} x^2\ln(-x^2) \langle g_s^2 G^2\rangle
  +\cdots,
  \label{eq:proplq}
\end{align}
where $\widetilde{x}\equiv i\hat{x}=i\gamma_\mu x^\mu$ and $G_{\mu\nu}^{ab}\equiv
G_{\mu\nu}^{n} T_{ab}^{n}=G_{\mu\nu}^{n}\lambda_{ab}^{n}/2$. There is also a
dangling-gluonic part \cite{Wang:2008vg,Wang:2011zzx}
\begin{align}
iS_{q,\mu\nu}^{ab,n}(x)
\equiv&\langle0|T\{q^a(x) g_s G_{\mu\nu}^n \bar{q}^b(0)\}|0\rangle \cr
=&-\frac{1}{2^6\times 3}\sigma_{\mu\nu}T_{ab}^n
   \langle g_s\bar{q}\sigma Gq\rangle \cr
 &+\frac{m_q}{2^8\times 3}
   (\sigma_{\mu\nu}\widetilde{x}+\widetilde{x}\sigma_{\mu\nu})T_{ab}^n
   \langle g_s\bar{q}\sigma Gq\rangle
  +\cdots ,
\label{eq:proplq2}
\end{align}
which is important for the condensate contributions from different quarks.

The heavy quark propagator in the coordinate space can be obtained from the Fourier
transformation of $iS_Q^{ab}(p)$ in \eqref{eq:prophQmom},
\begin{equation}
iS_Q^{ab}(x)
=\int\frac{d^4p}{(2\pi)^4} iS_Q^{ab}(p)\,e^{-ipx}.
\end{equation}
By using the Fourier transformation relation in $D$ dimensional Minkowski space
\begin{align}
i\int_\mathrm{M} \frac{d^D p}{(2\pi)^D}\frac{e^{-ipx}}{(p^2-m^2)^\nu}
=&\frac{m^{D-2\nu} r^{\nu-D/2} K_{\nu-D/2}(r)}
       {(-2)^{\nu-1} (2\pi)^{D/2}\Gamma(\nu)} ,
\end{align}
where $K_\nu(r)$ is the modified Bessel function of the second kind,
the heavy quark propagator in the coordinate space can be expressed as
\begin{align}
iS_Q^{ab}(x)
=&\frac{m^{3}\delta^{ab}}{(2\pi)^{2}}
  \left\{
  \widetilde{r}\,r^{-2} K_{2}(r)
 +r^{-1} K_{1}(r)
 \right\} \cr
&
 -\frac{m g_s G_{\mu\nu}^{ab}}{8(2\pi)^{2}}
  \left\{
  (\sigma^{\mu\nu}\widetilde{r}+\widetilde{r}\sigma^{\mu\nu}) r^{-1} K_{1}(r)
 +2\sigma^{\mu\nu} K_{0}(r)
 \right\} \cr
&
 -\frac{\delta^{ab} \langle g_s^2 G^2 \rangle}{576(2\pi)^{2}m}
  \left\{
  (\widetilde{r}-6)\,r^{1} K_{1}(r) +r^{2} K_{2}(r)
  \right\}
 +\cdots ,
 \label{eq:prophQ}
\end{align}
where~$\widetilde{r}\equiv m\widetilde{x}$, $r\equiv m\sqrt{-x^2}$ and
$\widetilde{r}^2=r^2$.

In the coordinate space, components of the light quark propagator
\eqref{eq:proplq} are of the form $(-x^2)^n$ or $\widetilde{x}(-x^2)^n$. Note
that the $(-x^2)^{n-2}\ln(-x^2)$ terms can be rewritten in the form of
$(-x^2)^{n-D/2}\Gamma(D/2-n)$. As for the heavy quark propagator
\eqref{eq:prophQ}, components are of the form $(-x^2)^n K_\nu(m\sqrt{-x^2})$
or $\widetilde{x}(-x^2)^n K_\nu(m\sqrt{-x^2})$. Naturally, the correlation
function \eqref{eq:piprops} can be expressed as $D$ dimensional Fourier
transformation of two Bessel functions.

%%}}}2
%%============================================================================
\subsection{Hypergeometric representation of the correlation function}%%{{{2
\label{sec:hyprep}

After some algebras of the Dirac gamma matrices, the Gell-Mann matrices and
the color indices \cite{Mertig:1990an}, the correlation function reads
\begin{equation}
\Pi^{\mu\nu}(q)\sim i\int_\mathrm{M} d^Dx\,e^{iq.x}
\{-g^{\mu\nu},x^\mu x^\nu\}
\sum_i \sqrt{-x^2}^{n_i}
K_{\nu_{i_1}}(m\sqrt{-x^2})K_{\nu_{i_2}}(m\sqrt{-x^2}) ,
\label{eq:pistruct}
\end{equation}
where $\{-g^{\mu\nu},x^\mu x^\nu\}$ are possible Lorentz structures and ``M''
indicates the above integral is calculated in the Minkowski space. The
angular integral of the Fourier transformation is trivial,
\begin{align}
i\int_\mathrm{M} d^Dx\,e^{iq.x}f(\sqrt{-x^2})=
(2\pi)^{D/2}Q^{1-D/2}
\int_0^\infty dx\,x^{D/2}J_{D/2-1}(Q x)f(x),
\label{eq:angular}
\end{align}
where $Q^2=-q^2$, $Q>0$, and $J_\nu(x)$ is the Bessel function of the first
kind.

Consequently, the correlation function $\Pi(q^2)$ turns into one dimensional
integrals of one Bessel $J$ and two Bessel $K$'s:
\begin{equation*}
\int_0^\infty dx\,x^{u-1}J_v(Q x)K_a(m x)K_b(m x).
\end{equation*}
This type of integrals can be worked out analytically with the help of
negative dimensional integration method (NDIM)
\cite{Suzuki:1997yz,Suzuki:1998qv, Anastasiou:1999ui} or its optimized
version, the method of brackets (MB)
\cite{Gonzalez:2007ry,Gonzalez:2008xm,Gonzalez:2011nq,Gonzalez:2010nm}. In
this calculation, Bessel functions are needed to be expressed in the form of
series. In detail,
\begin{equation}
J_\nu(x)=\sum_{m=0}^{\infty}\phi_m
  \frac{1}{\Gamma(m+\nu+1)}\left(\frac{x}{2}\right)^{2m+\nu} ,
\end{equation}
where $\phi_m=\tfrac{(-1)^m}{\Gamma(m+1)}$. Since $K_\nu(x)$ does not have a
single summation series representation, one needs to express $K_\nu(x)$ as a
definite integral
\begin{equation}
K_\nu(x)
=\frac{1}{2}\int_0^\infty dt\,t^{\nu-1}\,
  e^{-\frac{x}{2}\left(t+\frac{1}{t}\right)},
\end{equation}
or as a double summation series
\begin{align}
K_\nu(x)
&=\frac{1}{2}\sum_{n_1,n_2}\phi_{n_1,n_2}
  \left(\frac{x}{2}\right)^{n_1+n_2}
  \langle n_1-n_2+\nu \rangle ,
\end{align}
where $\phi_{n_1,n_2}=\phi_{n_1}\phi_{n_2}$ and a formal symbol $\langle a
\rangle$, the bracket, is introduced. This bracket is a short form of
the divergent integral
\begin{equation}
\int_0^\infty dx\,x^{a-1}\equiv\langle a \rangle.
\end{equation}

After a little algebra, one gets the key integral of the correlation function
calculation in coordinate space
\begin{align}
\hspace{\mathindent}&\hspace{-\mathindent}
\int_0^\infty dx\,x^{u-1}J_v(Q x)K_a(m x)K_b(m x) \cr
=&2^{u-3}m^{-u-v}Q^v
  \frac{\Gamma(A_{--})\Gamma(A_{+-})
        \Gamma(A_{-+})\Gamma(A_{++})}
       {\Gamma(1+v)\Gamma(u+v)} \cr
 &\times
  \HypergeometricPFQ{4}{3}%
  {A_{--},A_{+-},A_{-+},A_{++}}%
  {1+v,\tfrac{u+v}{2},\tfrac{1+u+v}{2}}%
  {\frac{-Q^2}{4m^2}}.
  \label{eq:JKK1}
\end{align}
Here the notation
\begin{equation*}
A_{\lambda_1\lambda_2}\equiv\frac{u+v+\lambda_1 a+\lambda_2 b}{2}
\end{equation*}
is used to keep the expression short. A similar integral with one Bessel $K$
has been given in the form of $\HypPFQ{2}{1}$ in Ref.~\cite{Groote:2005ay}, which
can be used to study hadrons with one heavy quark like $D$ and $B$ mesons as
well as $\Lambda_c$ and $\Lambda_b$ baryons. Since Feynman integrals can be
expressed as linear combinations of hypergeometric functions, and two-loop
self-energy integrals as well as two-loop sunset-type diagrams have
$\HypPFQ{4}{3}$ representations \cite{Davydychev:2003mv,Broadhurst:1993mw},
the above expression appears as expected.

Alternatively, the integral above can also be expressed as
\begin{align}
\hspace{\mathindent}&\hspace{-\mathindent}
\int_0^\infty dx\,x^{u-1}J_v(Q x)K_a(m x)K_b(m x) \cr
=&2^{u-3}m^{a+b}Q^{-u-a-b}
  \frac{\Gamma(-a)\Gamma(-b)\Gamma(\tfrac{u+v+a+b}{2})}
       {\Gamma(1-\tfrac{u-v+a+b}{2})} \cr
 &\times\HypergeometricPFQ{4}{3}%
  {\tfrac{1+a+b}{2},1+\tfrac{a+b}{2},
   \tfrac{u-v+a+b}{2},\tfrac{u+v+a+b}{2}}%
  {1+a,1+b,1+a+b}%
  {\frac{4m^2}{-Q^2}} \cr
 &{}+(a\to-a)+(b\to-b)+(a\to-a,\,b\to-b).
  \label{eq:JKK2}
\end{align}
These two representations \eqref{eq:JKK1} and \eqref{eq:JKK2} are simply
related to each other by analytic continuation of the hypergeometric functions.
\begin{align}
\HypergeometricPFQ{q+1}{q}{a_1,\ldots,a_{q+1}}{b_1,\ldots,b_q}{z}
=&\frac{\Gamma(b_1)\cdots\Gamma(b_q)}{\Gamma(a_1)\cdots\Gamma(a_{q+1})}
  \sum_{i=1}^{q+1}
  \frac{\Gamma(a_i)\prod_{j=1;j\neq i}^{q+1}\Gamma(a_j-a_i)}
                  {\prod_{j=1}^{q}\Gamma(b_j-a_i)} \cr
 &\times
  (-z)^{-a_i}
  \HypergeometricPFQ{q+1}{q}%
    {a_i,\{1+a_i-b_k\}_{k=1,\ldots,q}}%
        {\{1+a_i-a_k\}_{k=1,\ldots,q+1;k\neq i}}%
    {\frac{1}{z}},
  \label{eq:q1fqAC}
\end{align}
where $a_j-a_i\notin\mathbb{Z}$.
In the following calculation, \eqref{eq:JKK1} is used to derive the
hypergeometric representation of the correlation function because this
representation can be simply regularized by performing the coordinate space
integral in $D=4-2\epsilon$ dimension.

Using \eqref{eq:angular} and \eqref{eq:JKK1}, the hypergeometric
representation of the correlation function can be worked out as
\begin{align}
\PK_0=&
i\int_\mathrm{M} d^Dx\,e^{iq.x}
\sqrt{-x^2}^w K_a(m\sqrt{-x^2})K_b(m\sqrt{-x^2}) \cr
=&2^{D+w-2}\pi^{D/2}
  \frac{\Gamma(B_{--})\Gamma(B_{+-})
        \Gamma(B_{-+})\Gamma(B_{++})}
       {m^{D+w}
        \Gamma(\tfrac{D}{2})\Gamma(D+w)} \cr
 &\times\HypergeometricPFQ{4}{3}%
  {B_{--},B_{+-},
   B_{-+},B_{++}}%
  {\tfrac{D}{2},\tfrac{D+w}{2},\tfrac{D+w+1}{2}}%
  {z},
  \label{eq:pikk}
\end{align}
where $z=s/(4m^2)$ and $s=q^2=-Q^2$. Here the notation
\begin{equation*}
B_{\lambda_1\lambda_2}\equiv\frac{D+w+\lambda_1 a+\lambda_2 b}{2}
\end{equation*}
is used to keep the expression short. Note that $D$ appears in every Gamma
function and every parameter of the hypergeometric function, which
regularizes the divergences very well. It is helpful to use the dimensionless
variable $z$ as the argument of hypergeometric function. Consequently, all
dimensional constants like masses and condensates appear in the overall
factor. What's more, $z$ appears only in the hypergeometric function, which
makes it transparent that the correlation function has a discontinuity only for
$z\geq 1$ (or $s\geq 4m^2$).

In the spectral density calculation, the coordinate space integral is
performed in $D=4-2\epsilon$ dimension, while the propagators of light and
heavy quarks in $4$ dimension are employed. That is, $w$, $a$ and $b$ in
\eqref{eq:pikk} are all integers. The dimension parameter $D$ in the power of
constants like $2^{D+w-2}$, $\pi^{D/2}$ and $m^{D+w}$ can be set to $4$ safely,
because the $\epsilon$ related higher-order infinitesimal terms of the
constant coefficients will not lead to finite contributions to the spectral density.
Furthermore, $\HypPFQ{4}{3}$ will reduce to $\HypPFQ{3}{2}$ or even much
simpler functions, depending on the specific situation.

The correlation function contains Lorentz indices $\mu$ and $\nu$, which means
$\Pi_{\mu\nu}(x)$ contains $g_{\mu\nu}$ and $x_\mu x_\nu$ structures, as mentioned
in \eqref{eq:pistruct}. Terms
proportional to $g_{\mu\nu}$ can be directly calculated through
\eqref{eq:pikk}, and terms proportional to $x_\mu x_\nu$ can be calculated
through the differential form of \eqref{eq:pikk}.

The derivative of hypergeometric function is
\begin{equation*}
\frac{\partial}{\partial z}
\HypergeometricPFQ{q+1}{q}{\{a_i\}}{\{b_j\}}{z}
=\frac{\Pi\{a_i\}}{\Pi\{b_j\}}
  \HypergeometricPFQ{q+1}{q}{\{a_i+1\}}{\{b_j+1\}}{z} ,
\end{equation*}
where $\Pi\{a_i\}\equiv\Pi_{i=1}^{q+1} a_i$ and $\Pi\{b_j\}\equiv\Pi_{j=1}^{q} b_j$.
The double partial derivative of hypergeometric function is
\begin{align*}
-\frac{\partial^2}{\partial q_\mu \partial q_\nu}
\HypergeometricPFQ{4}{3}{\{a_i\}}{\{b_j\}}{\frac{q^2}{4m^2}}
=&-\frac{q^\mu q^\nu}{4m^4}
  \frac{\Pi \{a_i(a_i+1)\}}{\Pi \{b_j(b_j+1)\}}
  \HypergeometricPFQ{4}{3}{\{a_i+2\}}{\{b_j+2\}}{\frac{q^2}{4m^2}} \cr
 &-\frac{g^{\mu\nu}}{2m^2}
  \frac{\Pi \{a_i\}}{\Pi \{b_j\}}
  \HypergeometricPFQ{4}{3}{\{a_i+1\}}{\{b_j+1\}}{\frac{q^2}{4m^2}} .
\end{align*}
For the hadron state with $J^{PC}=1^{++}$, only the $(-g^{\mu\nu})$ part is needed in
the spectral density calculation. As a result, one gets
\begin{align}
\PK_2=&
i\int_\mathrm{M} d^Dx\,e^{iq.x}
x^\mu x^\nu
\sqrt{-x^2}^{w-2} K_a(m\sqrt{-x^2})K_b(m\sqrt{-x^2}) \cr
\to&
  2^{D+w-3}\pi^{D/2}
  \frac{\Gamma(B_{--})\Gamma(B_{+-})
        \Gamma(B_{-+})\Gamma(B_{++})}
       {m^{D+w}
        \Gamma(1+\tfrac{D}{2})\Gamma(D+w)} \cr
&\times
  \HypergeometricPFQ{4}{3}%
  {B_{--},B_{+-},
   B_{-+},B_{++}}%
  {1+\tfrac{D}{2},\tfrac{D+w}{2},\tfrac{D+w+1}{2}}%
  {z}.
\label{eq:pikk2}
\end{align}
Note that the factor $(-g^{\mu\nu})$ is omitted here for the sake of simplicity.

By using \eqref{eq:pikk} and \eqref{eq:pikk2}, one can work out the
hypergeometric representations of the correlation function $\Pi_1(q^2)$. The
parameters $\{a_i\}$ and $\{b_j\}$ of $\HypPFQ{q+1}{q}(\{a_i\},\{b_j\},z)$
type hypergeometric functions are linear functions of space-time dimension
$D=4-2\epsilon$, where $\epsilon$ regulates UV and/or IR divergences.

In fact, it is not too hard to express some integrals as generalized
hypergeometric functions, because it is trivial, to some extent, to transform
these integrals into summations. Nevertheless, if these hypergeometric functions can
not be easily handled, they are nothing more than short notations and the
problem remains unsolved. Fortunately, with the help of some new technologies,
the $\HypPFQ{q+1}{q}$ type hypergeometric functions can be effectively handled
in many ways.

To extract the spectral density form the hypergeometric representation of the
correlation function, two independent approaches are presented below. The
simple integral representation method in Sec.~\ref{sec:intrep} is easy-to-use,
where the spectral density will be expressed as one-dimensional integrals. In
Sec.~\ref{sec:hypexp}, the
$\epsilon$-expansion method of the hypergeometric functions can even express
the spectral density in terms of some commonly known functions and no integral
is needed.

%%}}}2
%%============================================================================
\subsection{Integral representation of hypergeometric function}%%{{{2
\label{sec:intrep}

Once the correlation function is expressed as hypergeometric functions like
in \eqref{eq:pikk} and \eqref{eq:pikk2}, one
can use the simple integral representations \cite{Broadhurst:1993mw} of
$\HypPFQ{2}{1}$ and $\HypPFQ{3}{2}$
\begin{align}
\HypergeometricPFQ{2}{1}{a_1,a_2}{b_1}{z}
=&\frac{\Gamma(b_1)}
       {\Gamma(a_2)\Gamma(b_1-a_2)}
  \int_1^\infty dt\,
  \frac{t^{a_1-b_1}(t-1)^{b_1-a_2-1}}{(t-z)^{a_1}},
  \label{eq:int2f1} \\
\HypergeometricPFQ{3}{2}{a_1,a_2,a_3}{b_1,b_2}{z}
=&\frac{\Gamma(b_1)\Gamma(b_2)}
       {\Gamma(a_2)\Gamma(a_3)\Gamma(b_1+b_2-a_2-a_3)} \cr
 &\times
  \int_1^\infty dt\,
  \frac{t^{a_1-b_1}(t-1)^{b_1+b_2-a_2-a_3-1}}{(t-z)^{a_1}} \cr
 &\quad\times
  \HypergeometricPFQ{2}{1}{b_2-a_2,b_2-a_3}{b_1+b_2-a_2-a_3}{1-t},
  \label{eq:int3f2}
\end{align}
to calculate the spectral density. It is worth noting that these integral
representations are peculiarly suitable for the discontinuity calculation. In
these representations, $z$ only appears in $(t-z)^{-a_1}$. Generally,
the discontinuity of $(t-z)^{-a_1}$ is~\cite{Groote:2005ay}
\begin{equation}
\frac{1}{2\pi i}\Disc(t-z)^{-a_1}
=\frac{(z-t)^{-a_1}\theta(z-t)}{\Gamma(a_1)\Gamma(1-a_1)}.
\label{eq:discpow}
\end{equation}
Note that for $a_1=n-\epsilon$, where $n$ is a positive integer, the expansion
\cite{Anastasiou:2003gr,Carter:2010hi}
\begin{equation}
z^{\epsilon-1}
=\frac{1}{\epsilon}\delta(z)
 +\sum_{m=0}^{\infty} \frac{\epsilon^m}{m!}\left[\frac{\log^m(z)}{z}\right]_+
\end{equation}
and its differential forms are used to calculate the discontinuities of the
above integrals. The ``plus'' distributions are defined as
\begin{equation}
\int dz\,f(z)\left[\frac{g(z)}{z}\right]_+=\int dz\,\frac{f(z)-f(0)}{z}g(z).
\end{equation}
For the high dimensional condensate parts of the correlation function,
$B_{\lambda_1\lambda_2}$ may be positive. The ``delta'' distribution is
more important than the ``plus'' one, because it will lead to
$\epsilon^{-1}\delta^{(n)}(z-t)$ type of contributions to the simple integral.
Note that the surface terms of the $\delta^{(n)}(z-t)$ related integrals may
contain finite contributions to the sum rules.

If the hypergeometric functions have already reduced to some simple functions
like $(z-1)^{-n}$, where $n$ is a positive integer, no integral representation
is needed and the discontinuity can be calculated as follows
\begin{equation}
\frac{1}{2\pi i}\Disc(-z)^{-n}=\frac{1}{(n-1)!}\delta^{(n-1)}(z).
\label{eq:discdelta}
\end{equation}

%%}}}2
%%============================================================================
\subsection{Epsilon expansion of hypergeometric function}%%{{{2
\label{sec:hypexp}

The spectral density can even be analytically worked out from the
hypergeometric representation of the correlation function. In the recent
decade, lots of algorithms and packages \cite{Moch:2001zr,Weinzierl:2002hv,
Huber:2005yg,Huber:2007dx, Kalmykov:2006hu,Kalmykov:2007pf, Bytev:2011ks} have
been developed to perform the $\epsilon$-expansion of hypergeometric
functions. Among them, \verb|HypExp|~\cite{Huber:2005yg,Huber:2007dx} provides
a systematic and easy-to-use approach to perform $\epsilon$-expansion of
$\HypPFQ{q+1}{q}$ type hypergeometric functions, whose parameters $\{a_i\}$
and $\{b_j\}$ are of the form $n+\alpha\epsilon$ or
$n+\tfrac{1}{2}+\alpha\epsilon$.

In \eqref{eq:pikk} and \eqref{eq:pikk2}, the overall factor of Gamma
functions has non-zero contribution from order $\mathcal{O}(\epsilon^{-n_1})$, while
hypergeometric function has non-zero discontinuity from order
$\mathcal{O}(\epsilon^{n_2})$. For the calculations in Sec.~\ref{sec:egapp},
it is interesting to see that $n_1=n_2$, and then the spectral density calculation
can be drastically simplified by only calculating $\epsilon^{-n_1}$ part of
the overall factor and $\epsilon^{n_1}$ part of the hypergeometric function.
Perhaps this favorable condition does not always exist, but it is worth checking
it beforehand.

After $\epsilon$-expansion, the hypergeometric functions (or the correlation
functions) are expressed in terms of commonly known functions and harmonic
polylogarithms (HPLs) \cite{Remiddi:1999ew,Maitre:2005uu}. The HPLs are of the form
\begin{equation}
H_{\cdots}\left(i\sqrt{-\frac{z}{z-1}}\right) ,
\end{equation}
where $\cdots$ denotes indices. Note that the argument $i\sqrt{-z/(z-1)}$ of
HPLs is not in the well-defined interval $(0,1)$ for $z>1$, one cannot
naively convert this kind of HPLs to commonly known functions for now.

Empirically, it is preferable to keep the HPLs unchanged before the
discontinuities are worked out because HPLs are compact functions. Otherwise,
one need to calculate the discontinuities of a large amount of common
functions. The discontinuities are calculated through
\begin{equation}
\rho(z)
=\frac{1}{2\pi i}\Disc\Pi(z)
=\lim_{\varepsilon\to0^+}\frac{\Pi(z+i\varepsilon)-\Pi(z-i\varepsilon)}{2\pi i}.
\label{eq:disceps}
\end{equation}
Particularly,
\begin{equation}
\left.
\sqrt{-\frac{z}{z-1}}
\right|_{z\pm i\varepsilon}
=\varepsilon\pm i\sqrt{\frac{z}{z-1}},
\quad\text{for } z>1,
\end{equation}
and now the argument of HPLs becomes $-\sqrt{z/(z-1)}+i\varepsilon$ or
$\sqrt{z/(z-1)}+i\varepsilon$, respectively.

Since the HPLs are well defined for arguments in the interval $(0,1)$
\cite{Maitre:2005uu,Maitre:2007kp,Huber:2007dx}, while the real parts
$-\sqrt{z/(z-1)}\in(-\infty,-1)$ and $\sqrt{z/(z-1)}\in(1,\infty)$, these HPLs
need to be analytically continued to the interval $(0,1)$ before converting
them into commonly known functions. The analytic
continuation sign $\delta$ in the small imaginary part $i\delta\varepsilon$ of
the argument is important. In the practical calculation, HPLs with argument
$-\sqrt{z/(z-1)}+i\varepsilon$ are analytically continued from $(-\infty, -1)$
to $(0,1)$ by the function \cite{Maitre:2005uu}
\begin{verbatim}
    HPLAnalyticContinuation[#, AnalyticContinuationSign -> 1,
        AnalyticContinuationRegion -> minftom1]&
\end{verbatim}
and HPLs with argument $\sqrt{z/(z-1)}+i\varepsilon$ are analytically
continued from $(1,\infty)$ to $(0,1)$ by the function
\begin{verbatim}
    HPLAnalyticContinuation[#, AnalyticContinuationSign -> 1,
        AnalyticContinuationRegion -> onetoinf]&
\end{verbatim}
After the analytic continuation, the argument of all HPLs becomes
$\sqrt{1-1/z}$, which is in the interval $(0,1)$ for $z>1$.

Now one can use \eqref{eq:disceps} to calculate the discontinuity of the
$\epsilon$-expanded expression. It is known from the practical calculations
that only a few HPLs (see \eqref{eq:rhohpl}) are needed to express the
spectral density in a neat form. Since the argument of all HPLs becomes
$\sqrt{1-1/z}\in(0,1)$ for $z>1$, it is safe to convert HPLs to commonly
known functions like logarithm or polylogarithm. Besides, HPLs can also be
numerically evaluated with high efficiency like the common functions
\cite{Maitre:2005uu}, and one can directly use HPLs in the numerical
calculation.

The discontinuity near the lower threshold $z=1$ is related to functions of
the form $(1-z)^{-a}$. If $a$ is a positive integer, one can use
\eqref{eq:discdelta} to calculate the discontinuity.
The balance parameter
\begin{equation}
\sigma=\sum_{j=1}^{q} b_j -\sum_{i=1}^{q+1} a_i
\label{eq:balance}
\end{equation}
is used to determine the $z\to1$ behavior of $\HypPFQ{q+1}{q}$ type
hypergeometric functions \cite{Buhring1992,Buhring2003}. If $\sigma<0$, the
singular parts of the hypergeometric function are of the form
$(1-z)^{\sigma+k}$, an additional regularization is needed to cancel such kind
of divergences. Such contributions might appear when the higher dimensional
condensates are taken into consideration.
From \eqref{eq:pikk} and \eqref{eq:pikk2}, it is easy to see that
$\sigma=(1-D)/2-w$ or $\sigma=(3-D)/2-w$. Since $w$ is an integer and $D$ will
be set to $4$ after the $\epsilon$-expansion, $\sigma$ cannot be a negative
integer. As a result, $(1-z)^{-n}$ will not appear in the correlation
function, and the spectral density will not contain Dirac delta function
related terms, either.

%%}}}2
%%============================================================================
%%}}}1
%%============================================================================
\section{Application}%%{{{1
\label{sec:egapp}

In this section, a concrete calculation is performed to show that the spectral
density of doubly heavy hadrons like hidden-charm tetraquark state
\cite{Matheus:2006xi} can be easily obtained by using the coordinate space
method.

The two-point correlation function \eqref{eq:pimunu} in the QCD spectral sum
rules can be decomposed into two parts as
\begin{align}
\Pi_{\mu\nu}(q)
&=\left(\frac{q_\mu q_\nu}{q^2}-g_{\mu\nu}\right)\Pi_1(q^2)
  +\frac{q_\mu q_\nu}{q^2}\Pi_0(q^2).
\end{align}
$\Pi_1(q^2)$ and $\Pi_0(q^2)$ have the quantum numbers of the spin 1 and 0
mesons, respectively. For the QCD spectral sum rule calculation of $X(3872)$
meson state with $J^{PC}=1^{++}$, only $\Pi_1(q^2)$ is of interest. The
correlation function and the spectral density are related by a dispersion
relation
\begin{equation}
\Pi_1(q^2)=\int_{s_<}^\infty ds\,\frac{\rho(s)}{s-q^2},
\end{equation}
where $s_<$ is the lower threshold and the spectral density is the discontinuity
of the correlation function
\begin{equation}
\rho(s)=\frac{1}{2\pi i}\Disc \Pi_1(s).
\end{equation}

The current of the $1^{++}$ tetraquark $[qc][\bar{q}\bar{c}]$ is given
by~\cite{Matheus:2006xi}
\begin{equation}
j_\Gamma=f_{ab;de}(q_a^T C\Gamma^A c_b)(\bar{q}_d \Gamma^B C \bar{c}_e^T),
\end{equation}
where $f_{ab;de}$ is the color structure factor, and the Dirac gamma matrix
$\Gamma$ is of the form $I$, $\gamma^\mu$, $\sigma^{\mu\nu}$,
$\gamma^5\gamma^\mu$ or $i\gamma^5$. Note that $\Gamma$'s satisfy the
transformation relation
\begin{equation}
\gamma^0\Gamma\gamma^0=\Gamma^\dag.
\end{equation}

The vacuum expectation value of the currents is
\begin{align}
\langle0|T\{j_\mu(x)j_\nu^\dag(0)\}|0\rangle
=&\langle0|T\{j_\mu(x)j_\nu^\dag(0)\}|0\rangle_1 \cr
 &+(\Gamma^{A}\leftrightarrow\Gamma^{B})
  +(\Gamma^{C}\leftrightarrow\Gamma^{D})
  +(\Gamma^{A}\leftrightarrow\Gamma^{B},
    \Gamma^{C}\leftrightarrow\Gamma^{D}).
\end{align}
The labeled gamma matrices $\Gamma^A$, $\Gamma^B$, $\Gamma^C$ and $\Gamma^D$ are
introduced to make the sources of the gamma matrices explicit. In the final
expression, these gamma matrices will be replaced by $i\gamma_{5}$,
$\gamma_\mu$, $i\gamma_{5}$ and $\gamma_\nu$, respectively, for the specific
spectral density calculation in Ref.~\cite{Matheus:2006xi}.

After contracting the quark fields in the correlation function with Wick
theorem, one part of the vacuum expectation values reads
\begin{align}
\langle0|T\{j_\mu(x)j_\nu^\dag(0)\}|0\rangle_1
=&\langle0|
  \Tr\{\Gamma^{A}[iS_{bb'}^{c}(x)]\Gamma^{C}(C[iS_{aa'}^{q}(x)]^T C^{-1})\} \cr
 &\times
  \Tr\{\Gamma^{B}(C[iS_{e'e}^{c}(-x)]^T C^{-1})\Gamma^{D}[iS_{d'd}^{q}(-x)]\}
  |0\rangle ,
\end{align}
where $iS_{aa'}^{q}(x)$ is the full quark propagator in coordinate space. It
is easy to obtain the other three parts by permutations of gamma matrices.

In the propagators \eqref{eq:proplq} and \eqref{eq:prophQ}, the gamma matrix
related structures are $1$, $\widetilde{x}$, $\sigma^{\mu\nu}$,
$\sigma^{\mu\nu}\widetilde{x}+\widetilde{x}\sigma^{\mu\nu}$. Their charge
conjugation transformations are
\begin{align}
C\{1,\widetilde{x},\sigma^{\mu\nu},
  \sigma^{\mu\nu}\widetilde{x}+\widetilde{x}\sigma^{\mu\nu}\}^{T}C^{-1}
=\{1,-\widetilde{x},-\sigma^{\mu\nu},
   \sigma^{\mu\nu}\widetilde{x}+\widetilde{x}\sigma^{\mu\nu}\}.
\end{align}
With the help of this identity, it is easy to get the charge conjugation of
the quark propagators. After some algebras of the gamma matrices, the
Gell-Mann matrices and the color indices \cite{Mertig:1990an}, the correlation
function can be obtained easily. For example, the perturbative part of the
correlator in coordinate space is
\begin{align}
\Pi_1^{\mathrm{pert},\mu\nu}(x)=
\frac{3m_c^4}{\pi^8}
\left(-g^{\mu\nu}(-x^2)^{-4} -2 x^\mu x^\nu(-x^2)^{-5}\right)
K_{-2}(m_c\sqrt{-x^2})^2 .
\label{eq:kkpert}
\end{align}
By using \eqref{eq:pikk} and \eqref{eq:pikk2} with $w=-8$ and $a=b=-2$,
one gets the $\Pi_1^{\mathrm{pert}}(z)$ in the form of hypergeometric functions.
Explicitly,
\begin{align}
\Pi_1^{\mathrm{pert}}(z)=
\frac{3m_c^8}{64\pi^6}\biggl\{
 &\frac{\Gamma(-\epsilon-4)\Gamma(-\epsilon-2)^2\Gamma(-\epsilon)}
       {\Gamma(2-\epsilon)\Gamma(-2\epsilon-4)}
  \HypergeometricPFQ{3}{2}{-\epsilon-4,-\epsilon-2,-\epsilon}{-\epsilon-\tfrac{3}{2},2-\epsilon}{z} \cr
-&\frac{\Gamma(-\epsilon-4)\Gamma(-\epsilon-2)^2\Gamma(-\epsilon)}
       {\Gamma(3-\epsilon)\Gamma(-2\epsilon-4)}
  \HypergeometricPFQ{3}{2}{-\epsilon-4,-\epsilon-2,-\epsilon}{-\epsilon-\tfrac{3}{2},3-\epsilon}{z}
  \biggr\} .
  \label{eq:pipert}
\end{align}
With the help of \eqref{eq:int3f2} and \eqref{eq:discpow}, one gets the
simple integral representation of the perturbative part of the spectral
density
\begin{align}
\rho^{\mathrm{pert}}(z)
=\frac{m_c^8}{\pi^6}\biggl\{
  \frac{32}{3465} &\int_1^z dt\,t^{3/2}(t-1)^{11/2}
  \HypergeometricPFQ{2}{1}{4,6}{\tfrac{13}{2}}{1-t} \cr
 -\frac{64}{45045}&\int_1^z dt\,t^{3/2}(t-1)^{13/2}
  \HypergeometricPFQ{2}{1}{5,7}{\tfrac{15}{2}}{1-t}
  \biggr\}.
  \label{eq:intpert}
\end{align}
Here $a_1=-\epsilon$ is chosen, which makes the $(z-t)^{-a_1}$ trivial when
$\epsilon$ expansion is performed. It is worth noting that one has the freedom
to chose a parameter from $\{a_i\}$ as $a_1$, for $\{a_i\}$ in $\HypPFQ{p}{q}$
are totally symmetric. With other choices of $a_1$, different but equivalent
integral representations will be obtained. Analogously, the other parts of the
spectral density can be expressed as simple integrals, too.

Furthermore, the spectral density can also be worked out analytically from
\eqref{eq:pipert} by using the $\epsilon$-expansion method in
Sec.~\ref{sec:hypexp}. Note that the $\epsilon$-expansion of the Gamma
functions in \eqref{eq:pipert} begins from $g_{-3}\epsilon^{-3}$, while
the non-vanishing discontinuity of the hypergeometric function $\HypPFQ{3}{2}$
begins from $f_3(z)\epsilon^{3}$.
Technically, it is sufficient to take $g_{-3}f_3(z)$ as the perturbative part
of the correlation function, which makes the spectral density calculation
drastically simplified. Explicitly, the perturbative part of the spectral
density is
\begin{align}
\rho^{\mathrm{pert}}(z)
=&\frac{m_c^8}{2^{10}\pi^6}
  \biggl\{
  \biggl[\frac{32z^4}{9}-\frac{1456z^3}{45}-\frac{776z^2}{15}+\frac{2624z}{45}
  +\frac{7}{18}+\frac{7}{12z}\biggr] V(z) \cr
 &\qquad
  +\biggl[48z^2-\frac{128z}{3}+5+\frac{7}{24z^2}\biggr] U(z)
  +12 T(z)
  \biggr\},
  \label{eq:hplpert}
\end{align}
where
\begin{equation}
V(z)=\sqrt{1-1/z},
\quad
U(z)=H_{+}(V(z)),
\quad
T(z)=H_{-,+}(V(z)).
\label{eq:rhohpl}
\end{equation}
The HPLs \cite{Remiddi:1999ew,Maitre:2005uu} $H_{+}(z)$ and $H_{-,+}(z)$ can
be converted to commonly known functions
\begin{align}
H_{+}(V(z))
 =&\log(1+V(z))-\log(1-V(z)),\\
H_{-,+}(V(z))
 =&\dilog\left(\tfrac{1-V(z)}{2}\right)
  -\dilog\left(\tfrac{1+V(z)}{2}\right)
  +\tfrac{1}{2}\log(4z)U(z),
\end{align}
where $\dilog(z)$ is the Euler dilogarithm.

In Ref.~\cite{Matheus:2006xi}, spectral density is calculated by the momentum
integral method. The perturbative part of the spectral density
is expressed by double integral of modified Feynman or Schwinger parameters
\begin{align}
\rho^{\mathrm{pert}}(s)
=&\frac{1}{2^{10}\pi^6}
  \int_{\alpha_<}^{\alpha_>}d\alpha
  \int_{\beta_<}^{\beta_>}d\beta
  \frac{1}{\alpha^3\beta^3}
  (1-\alpha-\beta)(1+\alpha+\beta)[(\alpha+\beta)m_c^2-\alpha\beta s]^4,
  \label{eq:intfp}
\end{align}
where $\alpha_<=(1-v)/2$, $\alpha_>=(1+v)/2$, $\beta_<=\alpha
m_c^2/(s\alpha-m_c^2)$ and $\beta_>=1-\alpha$, with $v=\sqrt{1-4m_c^2/s}$. This
double integral representation is obtained by calculating the discontinuity of
two-loop sunset type momentum integrals like \eqref{eq:mom2loop}.

$V(z)$ and $U(z)$ vary drastically near the threshold $z=1$, which means the
numerical computation of the double integral representation \eqref{eq:intfp}
might contain large relative errors near the threshold. In the QCD sum rules
approach, high energy part of the spectral density $\rho(z=s/(4m_c^2))$ is
damped by the factor $e^{-s\tau}$, and the large relative errors of the low
energy part of the spectral density might cause some noticeable errors to the
sum rules.
The double integrals are also time-consuming. Actually, three-fold numerical
integrals are involved in the sum rule calculation, if the $s$ integral is
taken into account. Numerically, \eqref{eq:intfp} is about hundreds of
times slower than \eqref{eq:hplpert}.

To make realistic uncertainty estimates of the results, a Monte-Carlo based
uncertainty analysis \cite{Leinweber:1995fn,Wang:2011zzx} is often used in the
QCD sum rule calculation. In this procedure, the entire phase space of QCD
input parameters like quark masses and condensates, is explored
simultaneously, and is mapped into uncertainties in the phenomenological
parameters. Since a Monte-Carlo based
uncertainty analysis need more than $100$ times of evaluation to get stable
uncertainties, it is not wise to repeatedly calculate the multi-dimensional
integrals. Technically, one can write the double integrals in a dimensionless
form by taking the dimensional parameters as the coefficients of these
integrals. Subsequently, the double integrals can be numerically evaluated
only once for a sequence of uniformly-spaced points $z_i=s_i/(4m_c^2)$, and
then the Newton-Cotes formula is used to perform the $z$ integration.

Comparing \eqref{eq:intpert} and \eqref{eq:hplpert} with
\eqref{eq:intfp}, one can find that the coordinate space method is more
suitable for spectral density and sum rule calculation, at least for the case
of two heavy quarks with equal masses.

In the same way, the explicit analytic expressions of the other parts of
the spectral density \cite{Matheus:2006xi} can be obtained as
\begin{align}
\rho^{m_q}(z)
=&\frac{m_q m_c^7}{2^8 \pi^6}
  \biggl\{
  \biggl[\frac{48z^3}{5}+\frac{617z^2}{15}-\frac{2087z}{30}
  -\frac{127}{24}-\frac{7}{16z}\biggr] V(z) \cr
 &\hspace{2.4em}
  -\biggl[28z^2-40z-3+\frac{5}{2z}+\frac{7}{32z^2}\biggr] U(z)
  -18 T(z)
  \biggr\} \cr
 &\hspace{-1.8em}
  +\frac{m_q m_c^4\langle\bar{q}q\rangle}{2^5 \pi^4}
  \biggl\{
  \biggl[2z^2-\frac{37z}{3}-\frac{47}{12}+\frac{1}{8z}\biggr] V(z)
  +\biggl[9-\frac{2}{z}+\frac{1}{16z^2}\biggr] U(z)
  \biggr\},\displaybreak[0]\\
\rho^{\langle\bar{q}q\rangle}(z)
=&\frac{m_c^5\langle\bar{q}q\rangle}{2^5\pi^4}
  \biggl\{
  -\biggl[\frac{14z^2}{3}+\frac{47z}{9}-\frac{41}{36}-\frac{5}{24z}\biggr]V(z)\cr
 &\hspace{3.6em}
 +\biggl[\frac{20z}{3}-3+\frac{1}{2z}+\frac{5}{48z^2}\biggr] U(z)
  \biggr\},\displaybreak[0]\\ 
\rho^{\langle G^2\rangle}(z)
=&\frac{m_c^4\langle g^2G^2\rangle}{2^9 3\pi^6}
  \biggl\{
  -\biggl[\frac{4z^2}{3}+\frac{5z}{3}-\frac{1}{2}\biggr] V(z)
  +\biggl[2z-1+\frac{1}{4z}\biggr] U(z)
  \biggr\},\\
\rho^{\mathrm{mix}}(z)
=&\frac{m_c^3\langle\bar{q}g\sigma Gq\rangle}{2^6\pi^4}
  \biggl\{
  \biggl[\frac{58z}{9}-\frac{7}{36}-\frac{7}{24z}\biggr] V(z)
  -\biggl[\frac{4z}{3}+\frac{3}{2}+\frac{7}{48z^2}\biggr] U(z)
  \biggr\},\\
\rho^{\langle\bar{q}q\rangle^2}(z)
=&\frac{m_c^2 \langle\bar{q}q\rangle^2}{12\pi^2} V(z).
\end{align}

Accordingly, the spectral densities of the doubly heavy meson state are
analytically worked out through the coordinate space method. Since no
multi-dimensional integral like \eqref{eq:intfp} is involved in the
spectral density and the analytic functions can be numerically calculated with
high efficiency, the errors of sum rules from spectral density functions are
minimized. Moreover, a Monte-Carlo based uncertainty analysis can be performed
to make realistic uncertainty estimates of the phenomenological parameters,
which will improve the predictive ability of QCD sum rules.

%%}}}1
%%============================================================================
\section{Numerical evaluation of spectral density}%%{{{1
\label{sec:numhyp}

As the hypergeometric functions might not be well known for the standard
community and the analytic reduction might be a little involved, a numerical
method is presented in this section for fast evaluation of the spectral
density from the $\epsilon$-regularized hypergeometric functions.
Consequently, no analytic $\epsilon$-expansion is needed, and the
regularization scheme is no longer restricted by the capabilities of
\verb|HypExp|~\cite{Huber:2007dx}.

Since the spectral density itself is finite \cite{Groote:1998ic} and
independent of the regularization parameter, an unorthodox but reasonable
regularization scheme can be used to obtain the spectral density. Formally,
the well regularized spectral density can be expressed in the form of
\begin{equation}
\rho(z,\epsilon)
=\sum_{n=1}^{m}\epsilon^{-n}\cdot0+\rho(z)+\mathcal{O}(\epsilon),
\label{eq:rhoeps}
\end{equation}
where $\rho(z,\epsilon)$ is the approximation of order $\mathcal{O}(\epsilon)$
of the true spectral density $\rho(z)$. If $\epsilon$ is numerically small
enough, $\rho(z,\epsilon)$ can be taken as $\rho(z)$. It is worth noting that
multiple precision computation is needed to get rid of roundoff errors from
the $\sum_{n=1}^{m}\epsilon^{-n}\cdot0$ part. Roughly, $-(m+1)\lg(\epsilon)$
digits of precision is enough to obtain the correct results.

Practically, \eqref{eq:pikk} and \eqref{eq:pikk2} with the regularization
parameter $D=4-2\epsilon$ are not good for numerical computation, because
$\HypPFQ{q+1}{q}(\{a_i\},\{b_j\},z)$ is evaluated through its analytic
continuation \eqref{eq:q1fqAC} for $z>1$. If the parameters
$a_j-a_i\in\mathbb{Z}$, the arguments of gamma functions as well as the
parameters of $\HypPFQ{q+1}{q}(\{a_i'\},\{b_j'\},1/z)$ may contain nonpositive
integers, and some spurious divergences may arise. Then, additional auxiliary
parameters are needed to regularize the expression, which may lead to a
complex calculation.

Using \eqref{eq:q1fqAC}, \eqref{eq:pikk} and \eqref{eq:pikk2}, it is
straightforward to get the hypergeometric representation of the correlation
function with argument $1/z$. In this section, this representation will be
used for numerical computation. For $z>1$, the hypergeometric functions with
argument $1/z$ have no branch cuts, and the discontinuities come only from the
$(-z)^a$ part. With the help of \eqref{eq:discpow}, one gets the
discontinuities of the integrals of $(-g^{\mu\nu})$ and $x^\mu x^\nu$
structures. Explicitly,
\begin{align}
\DK_0=&
\frac{1}{2\pi i}\Disc i\int_\mathrm{M} d^Dx\,e^{iq.x}
\sqrt{-x^2}^w K_a(m\sqrt{-x^2})K_b(m\sqrt{-x^2}) \cr
=&2^{-a-b-2}\pi^{D/2}m^{-D-w} z^{-\tfrac{D+w+a+b}{2}}
  \frac{\Gamma(-a)\Gamma(-b)}
       {\Gamma(1-\tfrac{D+w+a+b}{2})\Gamma(-\tfrac{w+a+b}{2})} \cr
 &\times\HypergeometricPFQ{4}{3}%
  {\tfrac{1+a+b}{2},\,1+\tfrac{a+b}{2},\,
   1+\tfrac{w+a+b}{2},\,\tfrac{D+w+a+b}{2}}%
  {1+a,\,1+b,\,1+a+b}%
  {\frac{1}{z}} \cr
 &+(a\to-a)+(b\to-b)+(a\to-a,\,b\to-b),
\label{eq:disckk}
\end{align}
and
\begin{align}
\DK_2=&
\frac{1}{2\pi i}\Disc i\int_\mathrm{M} d^Dx\,e^{iq.x}
x^\mu x^\nu
\sqrt{-x^2}^{w-2} K_a(m\sqrt{-x^2})K_b(m\sqrt{-x^2}) \cr
\to&
  2^{-a-b-3}\pi^{D/2}m^{-D-w} z^{-\tfrac{D+w+a+b}{2}}
  \frac{\Gamma(-a)\Gamma(-b)}
       {\Gamma(1-\tfrac{D+w+a+b}{2})\Gamma(1-\tfrac{w+a+b}{2})} \cr
 &\times\HypergeometricPFQ{4}{3}%
  {\tfrac{1+a+b}{2},\,1+\tfrac{a+b}{2},\,
   \tfrac{w+a+b}{2},\,\tfrac{D+w+a+b}{2}}%
  {1+a,\,1+b,\,1+a+b}%
  {\frac{1}{z}} \cr
 &+(a\to-a)+(b\to-b)+(a\to-a,\,b\to-b).
\label{eq:disckk2}
\end{align}
Note that the factor $(-g^{\mu\nu})$ is omitted here for the sake of simplicity.
These two functions $\DK_0(D,w,a,b,m,z)$ and $\DK_2(D,w,a,b,m,z)$ can be well
regularized by an unorthodox regularization scheme: $D=4$,
$a=a_0+\alpha\epsilon$ and $b=b_0+\beta\epsilon$, with $\alpha\neq0$,
$\beta\neq0$ and $\alpha\pm\beta\neq0$. It is preferable to use irrational
$\alpha$ and $\beta$ for practical calculation.

Accordingly, it is easy to write out the spectral density from
$\DK_0(D,w,a,b,m,z)$ and $\DK_2(D,w,a,b,m,z)$. It is better to show the
utility of this algorithm by a concrete example. From \eqref{eq:kkpert}, one
can write out the spectral density
\begin{equation}
\rho^{\mathrm{pert}}(z,\epsilon)
=\frac{3m_c^4}{\pi^8}\bigl\{
  \DK_0(D,w,a,b,m_c,z)-2\DK_2(D,w,a,b,m_c,z)
\bigr\}
\end{equation}
with $D=4$, $w=-8$, $a=-2+\epsilon$ and $b=-2+3\epsilon$. Due to the
$\Gamma(-n+\alpha\epsilon)$ terms, a high precision computation is needed to
get rid of roundoff errors. Fortunately, the hypergeometric function can be
evaluated to arbitrary numerical precision by computation systems like the
Python library
\verb|mpmath|~\cite{mpmath} or Mathematica. Specifically, some numerical
results are listed in Table~\ref{tab:numhyp}. Note that the charm quark mass
$m_c$ is set to $1$, and the working precision is set to $32$ digits for
$\epsilon=10^{-8}$. The exact value of $\rho^{\mathrm{pert}}(z)$ can be
evaluate by using \eqref{eq:hplpert} or \eqref{eq:intfp}.
\begin{table}[hbt]
\centering
\caption{Numerical evaluation of spectral density $\rho^{\mathrm{pert}}(z)$
with $m_c=1$.}
\label{tab:numhyp}
\begin{tabular*}{\linewidth}{@{\extracolsep{\fill}}cccccc}
\toprule
& $z=1.1$ & $z=2$ & $z=5$ & $z=10$ & $z=20$ \\
& $(\times10^{-13})$ & $(\times10^{-7})$ & $(\times10^{-4})$
& $(\times10^{-2})$  & $(\times10^{-1})$ \\
\midrule
$\epsilon=10^{-4}$&3.847514018&3.484073813&3.790001592&1.489432452&3.687774818\\
$\epsilon=10^{-6}$&3.847553295&3.484354676&3.790789489&1.489904972&3.689380738\\
$\epsilon=10^{-8}$&3.847553688&3.484357485&3.790797370&1.489909698&3.689396801\\
Exact             &3.847553692&3.484357513&3.790797449&1.489909746&3.689396963\\
\bottomrule
\end{tabular*}
\end{table}

From Table~\ref{tab:numhyp}, it is easy to see that one can use a finite small
parameter to regularize the spectral density and obtain the numerical result
to a desired precision. Several different input values of $\epsilon$ parameter
are used to show the numerical stability of the algorithm. As $\epsilon$
decreases to zero, the numerical result steadily approaches the exact value,
yet more digits of working precision is needed. Therefore, it is
preferable to use a fairly small $\epsilon$ to save the computation time.

The numerical method can also be applied to the calculations of more
complicated hypergeometric functions. 

%%}}}1
%%============================================================================
\section{Conclusion}%%{{{1
\label{sec:con}

In this paper, a systematic and easy-to-use method is developed to
calculate directly the doubly heavy hadron spectral density in the coordinate space.
The correlation function is expressed in terms of generalized
hypergeometric functions, and then the spectral density is obtained through
two independent approaches: the simple integral representation method and the
$\epsilon$-expansion method, respectively. It is found that the spectral
density of doubly heavy hadrons can be analytically expressed through commonly
known functions. After that, a concrete calculation is performed to show that
the spectral density of doubly heavy hadrons can be easily obtained by using
the coordinate space method. This method can drastically simplify the QCD
spectral sum rule calculation of the $\{ccX\}$ and $\{bbX\}$ systems.

An instructive numerical method is developed to evaluate the
$\epsilon$-regularized spectral density directly. This method can also be
applied to other calculations where expressions are in the form of regularized
(hypergeometric) functions. In particular, the numerical computation can be
used to handle problems that are beyond the reach of the analytic
$\epsilon$-expansion method.

Since the spectral density is expressed in terms of simple functions, there is
no noticeable errors from multi-dimensional numerical integrals. A Monte-Carlo
based uncertainty analysis \cite{Leinweber:1995fn,Wang:2011zzx} can be used to
make realistic uncertainty estimates of the phenomenological parameters. Such
analysis can quantitatively improve the predictive ability of QCD sum
rules. Moreover, this method can also serve to be an important cross-check of
the widely used momentum representation method.

Furthermore, for a hadron state with two different massive quarks, the
correlation function can be expressed in the form of Appell $F_4$ functions of
two variables, as is shown in \ref{sec:appell}. This type of spectral density
can be used to study the $B_c$-like $\{cbX\}$ structures \cite{Zhang:2009vs,
Wang:2012kw,Albuquerque:2012rq}. It is of great worth to work out the spectral
density from the Appell $F_4$ functions. The calculations will be more
complicated because the Appell $F_4$ is intricate and not widely studied like
the generalized hypergeometric functions $\HypPFQ{p}{q}$. The method can also
be extended to $\{QQQX\}$ and $\{QQQ'X\}$ systems. These issues will be
discussed in detail in a future publication.

%%}}}1
%%============================================================================
\section*{Acknowledgments}%%{{{1
Z.W.H.~thanks Ze-kun Guo for helpful discussions.
This work is supported by the National Natural Science Foundation of China
under Grant No.~10775105, BEPC National Laboratory Project R\&D and BES
Collaboration Research Foundation, and the project of Wuhan University of
China under the Grant No.~201103013 and 9yw201115.

%%}}}1
%%============================================================================
\appendix
%%============================================================================
\section{Two massive quarks with different masses}%%{{{1
\label{sec:appell}

In the case of two heavy quarks with different masses, one can also use the
method of brackets
\cite{Gonzalez:2007ry,Gonzalez:2008xm,Gonzalez:2011nq,Gonzalez:2010nm}
to calculate the integral of Bessel functions. Explicitly,
this type of integral can be expressed in the form of Appell $F_4$.
\begin{align}
\hspace{\mathindent}&\hspace{-\mathindent}
\int_0^\infty dx\,x^{u-1}J_v(Qx)K_a(m_1 x)K_b(m_2 x) \cr
=&2^{u-3}m_1^{-u-v+b}m_2^{-b}Q^{v}
  \frac{\Gamma(b)\Gamma(A_{+-})\Gamma(A_{--})}
       {\Gamma(1+v)}
  \AppellF{4}{A_{+-};A_{--}}{1+v,1-b}%
             {\frac{-Q^2}{m_1^2},\frac{m_2^2}{m_1^2}} \cr
 &+(b\to-b).
\end{align}
Appell $F_4$ is defined as
\begin{align}
\AppellF{4}{\alpha;\beta}{\gamma_1,\gamma_2}{x,y}
&=\sum_{m,n=0}^{\infty}
  \frac{(\alpha)_{m+n}(\beta)_{m+n}}{m!n!(\gamma_1)_m(\gamma_2)_n}
  x^m y^n .
\end{align}

The above integral can also be expressed as Appell $F_4$ with arguments
($-Q^2/m_2^2$, $m_1^2/m_2^2$) or ($m_1^2/(-Q^2)$, $m_2^2/(-Q^2)$). These three
representations are related to each other by analytic continuation. If
$m_1=m_2=m$, $F_4$ turns into $\HypPFQ{4}{3}$.

With the help of \verb|XSummer| \cite{Moch:2005uc} and/or \verb|HYPERDIRE|
\cite{Bytev:2011ks}, $F_4$ may be reduced to commonly known functions and
HPLs. If the analytic $\epsilon$-expansion procedure fails, one can use
the method in Sec.~\ref{sec:numhyp} to evaluate the spectral density
numerically.

%%}}}1
%%============================================================================
\bibliographystyle{utcaps}
\bibliography{specfunc}
%%============================================================================
\end{document}